\documentclass{article}[12pt]
\usepackage{graphicx}

\title{Principal component analysis -–
an efficient tool for variable stars diagnostics}

\author{ZDEN\v{E}K MIKUL\'{A}\v{S}EK$^{1,\,2}$\\$^1$Institute of Theoretical
Physics and Astrophysics,\\ Masaryk University,  Kotl\'a\v{r}sk\'a
2, Brno, Czech Republic\\$^2$Observatory and Planetarium of
J.Palisa, Ostrava, Czech Republic\\
E-mail: mikulas@physics.muni.cz}

\begin{document}

\maketitle

\abstract{We present two diagnostic methods based on ideas of
Principal Component Analysis and demonstrate their efficiency for
sophisticated processing of multicolour photometric observations
of
variable objects.\\

}

\section{Introduction}

In the last few decades it has boomed volume and common access to
high-quality variable stars observational data, however the standard
used methods of data processing and interpretation have lagged
behind this progress. One of the step how to overtake this disaccord
is the consequent application of \textbf{P}rincipal
\textbf{C}omponent \textbf{A}nalysis (PCA) combined with the
\textbf{R}obust \textbf{R}egression (RR), factor analysis, wavelet
analysis and other sophisticated approaches to the treatment of
variable stars observations.

The commonly used method for the treatment of astrophysical data is
simple (unweighted) \textbf{L}east \textbf{S}quare \textbf{M}ethod
(LSM). As these data usually suffer from outliers and very different
quality, the method yields questionable and misleading results.
Robust regression as an adequate alternative of the standard LSM is
used only seldom, LSM weights, if introduced at all, are often used
unknowledgeable.

\section{Standard and Weighted PCA}

Principal component analysis is one of the oldest and the most
elaborated method of the treatment of statistical data. PCA can be
used to simplify a data-set without loss of information. It is a
linear transformation that chooses a new coordinate system such
that the greatest variance corresponds to the first axis, the
residuals then to second one, etc. PCA is simple, straightforward,
it does not need any model. It diminishes the number of
uncorrelated parameters necessary for the description of the
data-set, helps to reveal hidden relationships and effectively
suppresses noise. For more details see e.g. \cite{Jack2003} or
\cite{Jol2004}.

Presently, it is profusely used namely in image techniques,
politics, criminal science, sociology and other human sciences,
however, in astronomy is almost unknown. We shall demonstrate how
to apply the standard PCA on routine tasks of the variable stars
observations processing.

Let we have $p$ photometric measurements obtained in $q$
photometric colours, which we can arrange in the form of $p$ row
vectors with $q$ components:
$\{\mathbf{y_1,y_2,\ldots,y_\mathit{p}}\}$,
$\mathbf{y}_i=[y_{i1},y_{i2},\ldots y_{iq}]$, or into the
\textit{p}$\times$\textit{q} matrix $\mathbf{Y}$. Each measurement
can be then described as a point in the $q$-D space, all $p$
observations represent the ``cloud" of points, whose global
characteristics we will study by means of the standard PCA.

If we want to use PCA as effective as possible we shall linearly
transform components of these data vectors into new variables
$\{\mathbf{z_1,z_2,\ldots,z_\mathit{p}}\}$:

\begin{equation}
z_{ij}=\frac{y_{ij}-\overline{y_j}}{\overline{s_j}},\label{Eq1}
\end{equation}
where $\overline{y_j}$ is the mean value of the $j$-th components
($j$-th colour), $\overline{s_j}$ is the estimate of the mean
(typical) error (uncertainty) of the $j$-th component. The purpose
of this transformation is to identify the middle of the data cloud
of observations with the origin of the new system of coordinates and
to equalize all coordinates among them. The PCA here implicitly
hypothesizes that at least the ratios among errors of measurements
in various colours are roughly constant. ``Errorboxes" of particular
measurements in $q$-D space should have the form of spheres of the
unit radius.

The standard PCA can be easily extended to \textbf{W}eighted
\textbf{P}rincipal \textbf{C}omponent \textbf{A}nalysis (WPCA)
introducing weights of individual data vectors. The weight of that
is put to be inversely proportional to the square of
$\varepsilon_i$: $w_i\sim\varepsilon_i^{-2}$, where $\varepsilon_i$
is the expected uncertainty of a component of the $i$-th data vector
$\mathbf{z}_i$. Let $\mathbf{w}=[w_1,w_2,\ldots,w_\mathit{p}]$ is a
vector describing weights of individual data vectors, the diagonal
matrix of weights $\mathbf{W}$ of size \textit{p}$\times$\textit{p}
is defined: $\mathbf{W}=\mathrm{diag}(\mathbf{w})$. In our $q$-D
representation it corresponds to the permission that errorspheres of
individual sets of multicolour measurements may have various
effective radii, proportional to $\varepsilon_i$. The standard PCA
is then the special case for WPCA with equal weights, $\mathbf{W\sim
I}_p$.

The above mentioned PCA linear transformation of a vector
$\mathbf{z}$ to a smoothed vector $\mathbf{z_s}$ by actuation of the
smoothing \textit{q}$\times$\textit{q} matrix
$\widehat{\mathbf{A}}$, can be written as:

\begin{equation}
\mathbf{z_s}=\mathbf{z\,\widehat{A}}=\mathbf{z\,(A\,{A\!}^T}),\ \
\mathbf{y_s}=[\,z_{\mathrm{s1}}\,\overline{s_1}+\overline{y_1},\,
\ldots z_{\mathrm{s}\mathit{q}}\,\overline{s_q}+\overline{y_q}\,],
\label{Eq2}
\end{equation}
where $\mathbf{A}$ is the \textit{q}$\times$\textit{r} matrix
consisting of $r$ columns of normalized eigenvectors $\mathbf{a}_i$
of the symmetric definite \textit{q}$\times$\textit{q} matrix
$\mathbf{Z^TW\,Z}$, where $\mathbf{Z}$ is the
\textit{q}$\times$\textit{p} date matrix: $\mathbf{Z=[z_1;z_2;\dots
z_\mathit{q}]}$. As it follows from the definition, each eigenvector
$\mathbf{a}_i$ together with the corresponding eigenvalue
$\lambda_i$ shall obey the relation:
\begin{equation}
\mathbf{(Z^TW\,Z)\,a}_i=\mathbf{a}_i\,\lambda_i.
      \label{Eq3}
\end{equation}

It can be proven that for the \textit{q}$\times$\textit{q} matrix
$\mathbf{Z^TW\,Z}$ just $q$ eigenvalues and $q$ normalized
eigenvectors exist. All the set of $q$ eigenvectors forms an
orthonormal vector base. Let we order eigenvectors according to
their eigenvalues from the largest to the smallest ones into the
sequence $\{\mathbf{a}_1,\ldots\mathbf{a}_q\}$. Now we take the
first $r$ ($r\leq q$) eigenvectors and connect them into the matrix
$\mathbf{A}=[\mathbf{a}_1,\ldots\mathbf{a}_r]$. Their eigenvalues
then create the diagonal of the \textit{r}$\times$\textit{r}
diagonal matrix
$\mathbf{\Lambda}=\mathrm{diag}([\lambda_1,\ldots\lambda_r])$:

\begin{equation}
\mathbf{(Z^TW\,Z)\,A}=\mathbf{A\,\Lambda}
      ;\ \  \mathbf{a}_i\cdot \mathbf{a}_j=\delta_{ij},\ \ \Rightarrow
      \ \ \mathbf{{A\!}^TA}=\mathbf{I}, \label{Eq4}
\end{equation}
where $\delta_{ij}$ is the discrete version of the Kronecker delta
function, $\mathbf{I}$ is the \textit{r}$\times$\textit{r} identity
matrix. Vectors $\{\mathbf{a}_1,\mathbf{a}_2,\ldots,\mathbf{a}_r\}$
contained in $\mathbf{A}$ represent orthonormal vector base of the
$r$-D subspace plunged into the $q$-D space. The arranged set of
scalar products of a vector $\mathbf{z}$ and vectors
$\{\mathbf{a}_i\}$: $\{k_1,k_2,\ldots,k_r\}$, where
$k_i=\mathbf{z\cdot a}_i$, define a vector $\mathbf{k}$:

\begin{equation}
\mathbf{k}=\mathbf{z\,A};\ \
\mathbf{z_s}=\mathbf{k\,{A\!}^T}=\mathbf{z\,A\,{A\!}^T},\ \
\mathbf{Z_s}=\mathbf{K\,{A\!}^T}=\mathbf{Z\,A\,{A\!}^T}\label{Eq5}
\end{equation}
We can introduce the \textit{q}$\times$\textit{r} matrix
$\mathbf{K}$, $\mathbf{K}=[\mathbf{k}_1;\ldots \mathbf{k}_q]$.
Assuming the Eq.\,(\ref{Eq4}) we can write:

\begin{equation}
\mathbf{K}=\mathbf{Z}\,\mathbf{A}\ \ \Rightarrow\ \
\mathbf{{A\!}^T}\mathbf{(Z^TW\,Z)}\,\mathbf{A}=\mathbf{K^TW\,K}=(\mathbf{{A\!}^T}
\mathbf{A})\,\mathbf{\Lambda}=\mathbf{\Lambda}.
 \label{Eq6}
\end{equation}
The equation (\ref{Eq6}) shows us that eigenvalues correspond to sum
of weighted variance of the projections of all vectors
{$\mathbf{z}_i$} and gives us the reason why we should confine
ourselves only to such components for which their eigenvalues are
sufficiently large - others do not content any true information,
they represent only a noise and so could be trimmed.

\begin{figure}[h]
\begin{center}
\includegraphics[width=0.68\textwidth]{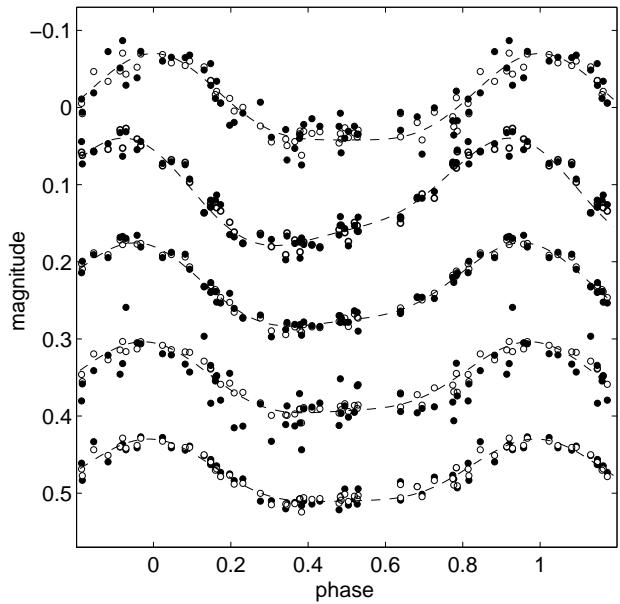}
\caption {Smoothing of ``observational" data (full dots) by the
standard PCA method. Smoothed points are denoted by circles, dashed
line represents the original unnoised light curves in five synthetic
colours.}\label{Fig1}
\end{center}
\end{figure}

The application of PCA and WPCA should help us to find the number of
parameters essential for the description of variability (number of
mechanisms of variability in action), it enables us to examine
relative quality of observations in multicolour measurements. Though
we do not know $s_j$ of individual colours exactly, we could improve
them very quickly using an iterative circle. The convergence of this
process is pretty good, because the results are as a rule
insensitive to the $s_j$ used.

Above mentioned methods help us namely in the preliminary processing
of observational data, when we want to reach an orientation in the
nature of variability of studied objects, possible relationships
among measured quantities and their quality. All these information
we can gain without using any physical model and time dependent
smoothing, what can strongly embarrass finding a priori unexpected
types of variability (rapid variations, trends etc.).

We demonstrate the PCA treatment of artificial photometric data (50
observations in 5 colours) simulating light variability of a
rotating CP star with two differently coloured photometric spots.
The ``observed" points and smoothed points with suppressed noise for
$r=2$ for individual colours are displayed in the phase diagram in
the Fig.~1. The treatment have not taken into its consideration the
phase information.

PCA methods similarly as LSM suffer from outliers which are quite
common in astrophysical data. Introducing of weights in PCA enables
us to eliminate their influence by means of an iterative process
adjusting individual weights of entering data (see e.g. the appendix
of \cite{MikRR2003}).

\section{Advanced PCA}

The extent of applicability of the standard and weighted PCA methods
is rather limited as they are demanding to the completeness and
homogeneity of input data. These confinements were obviously one of
decisive reasons why the PCA techniques remain beyond the scope of
the majority of observing astronomers.

Since 2000 we have developed a qualitatively new method synthesizing
weighted PCA and robust regression. We will denote it as Advanced
PCA (APCA). The versatility of APCA proves to be quite broad, it was
used several times, see e.g. \cite{MikGr2005}, \cite{MikKr2007},
\cite{MikZv2004}, however, it has not been fully described up to
now. We will briefly present only the method, without its derivation
and strict mathematical proving of lemmas or statements.

\subsection{Vector description of light curves}

Let the course of a light curve is described by means of preselected
model the parameters of which are determined by standard regression
methods, as LSM with weights or its modification eliminating the
influence of outliers. It is advisable to use a linear model so that
the course of a light curve in the certain colour $c$, $m_c(t)$
would be described by linear combination of the ensemble of $q$ so
called \emph{elementary} functions {$f_1(t),f_2(t),\ldots f_q(t)$}
defining the time dependence by the column vector of the length $q$:
$\mathbf{f}(t)=[f_1(t); f_2(t);\ldots f_q(t)]$ by the relation:
\begin{equation}
\Delta m_c(t)=m_c(t)-\overline{m_c}=\sum_{i=1}^{q}\,
y_{ci}\,f_i(t)=\mathbf{y}_c\,\mathbf{f}(t), \label{Eq7}
\end{equation}
where $y_{ci}$ are components of a row vector $\mathbf{y}_c$,
$\overline{m_c}$ is the mean magnitude in the colour $c$. The
components of the vector are found from the observational data by
standard regression procedures (weighted LSM, robust regression).

We should be very particular about the choice of the base of
elementary functions {$f_1(t),f_2(t),\ldots f_q(t)$}. The functions
should be selected so they enable us to express courses of all
studied light curves of the object with sufficient accuracy. It is
advisable for many reasons (avoiding of problems with
multicollinearity, equality of the uncertainty of components of the
vector $\mathbf{y}_c$) to opt elementary functions so that they
would form the base quasiorthonormal on the set of data, what means:

\begin{equation}
\overline{f_i^2}\cong\overline{f_j^2}\cong\overline{f^2};\ \ \ \
\overline{f_i\,f_j}\ll\overline{f^2}\ \ \mathrm {for}\ i\neq j.
\label{Eq8}
\end{equation}
In the case the set of elementary functions does not obey above
given conditions it is trivial to transform the system into
orthonormal one by means of standard Gram-Schmidt's
orthonormalization procedure.

Assuming observational data be distributed more or less uniformly
over the observational interval it is recommended to use Legendre
polynomials orthonormal on the interval $\langle-1;1\rangle$. If the
object is periodically variable then the condition of
quasiorthogonality fulfill any combination of harmonic polynomials
$\sin(2k\pi t/P),\ \cos(2k\pi t/P)$, $k=1,\,2,\ldots$;
$\overline{f^2}=1/2$.

If the functions of the linear regression model are quasiorthonormal
it is valid that uncertainties of particular components
$\varepsilon_c$ of the vector $\mathbf{y}_c$ are the same:
\begin{equation}
 \varepsilon_c\cong\frac{s_c}{\sqrt{N_c\,\overline{f^2}}};\ \ \ \
 w_c\sim\varepsilon_c^{-2}\sim\frac{N_c}{s_c^2}, \label{Eq9}
\end{equation}
where $s_c$ is the standard deviation of the light curve fit, $N_c$
is the number of observations in the particular colour used for the
light curve fit. The weight of the corresponding vector of light
curve in the $c$-colour then will be proportional to the
$\varepsilon_c^{-2}$.

The whole set of vectors describing the light curves in all $p$
colours can be arranged into the \textit{p}$\times$\textit{q} matrix
$\mathbf{Y}$, with the weights described by the
\textit{p}$\times$\textit{p} diagonal matrix $\mathbf{W}$.

\subsection{Advanced PCA. Reducing free parameters. Usage of APCA}

Let we permit that the variable part of light curves in all colours
$\Delta m_c$ can be sufficiently accurately approximated by linear
combination only $r$, ($r<q$) normalized orthogonal
(\emph{principal}) functions $\varphi_j(t)$ determined by linear
combination of all $q$ elementary function $f_i(t)$ with
coefficients forming the \textit{q}$\times$\textit{r} matrix
$\mathbf{B}$:

\begin{equation}
\varphi_j(t)=\sum_{i=1}^{q}\,b_{ij}\,f_i(t)=\mathbf{b}_j\,\mathbf{f}(t).
\label{Eq10}
\end{equation}
\begin{equation}
\Delta m_c(t)=\sum_{j=1}^{r}\,k_{cj}\,\varphi_j(t)=
\sum_{j=1}^{r}\,k_{cj}\mathbf{b}_j\,\mathbf{f}(t)=
\mathbf{k_\mathit{c}\,{B\!}^T\,f}(t)=\mathbf{y_{s\!\mathit{c}}}\,\mathbf{f}(t),
\label{Eq11}
\end{equation}
\begin{equation}
\mathbf{y_{s\!\mathit{c}}}=\mathbf{k_\mathit{c}\,{B\!}^T},\label{Eq12}
\end{equation}
where $\mathbf{b}_j$ is the normalized vector of the $j$-th
principal function and $j$-th column of the matrix $\mathbf{B}$. The
row vector $\mathbf{k}_c$ (1$\times$\emph{r}) represents
semiamplitude components of the light curve in colour $c$ versus $r$
principal functions $\{\varphi_1(t),\ldots\varphi_r(t)\}$ and the
1$\times$\emph{q} vector $\mathbf{y_{s\!\mathit{c}}}$ contains
parameters of the APCA smoothed light curve in the $c$ colour.

Further we will assume that the vector base
$\{\mathbf{b}_1,\ldots\mathbf{b}_r\}$ is orthonormal, then:
\begin{equation}
\mathbf{b}_i\,\mathbf{b}_j=\delta_{ij},\ \ \mathbf{{B\!}^TB=I}.
\label{Eq13}
\end{equation}

Minimizing the scalar quantity $S(\mathbf{B,k}_c)$ defined as the
the sum of weighted variances of differences
$\Delta\mathbf{y}_c=\mathbf{y}_c -\mathbf{y_{s\!\mathit{c}}}$:
\begin{equation}
S(\mathbf{B,k}_c)=\sum_{c=1}^{p}\Delta\mathbf{y}_c\,\Delta\mathbf{y}_c^\mathbf{T}\,w_c=
\sum_{c=1}^{p}(\mathbf{y_c\,y_c^{T}-k_c\,k_c^{T}})\,w_c ;\ \ \
\mathrm{grad}S=\mathbf{0}, \label{Eq14}
\end{equation}
we arrive after some algebra at the following conclusions:
\begin{equation}
\mathbf{K=YB},\ \ \mathbf{(Y^TW\,Y)\,B=B\,(K^TW\,K)=B\Lambda}.\ \
 \label{Eq15}
\end{equation}
\begin{equation}
\mathbf{Y_s=Y\,\widehat{B}=Y(B\,{B\!}^T)}. \label{Eq16}
\end{equation}

The results in equation (\ref{Eq15}) and  (\ref{Eq16}) are formally
identical with (\ref{Eq4}-\ref{Eq6}), so we can conclude that the
matrix $\mathbf{B}$ contains $r$ column vectors which are
eigenvectors of the matrix $\mathbf{Y^TW\,Y}$ corresponding to its
first $r$ largest eigenvalues. The smoothing matrix
$\mathbf{\widehat{B}}$ is defined identically, as
$\mathbf{\widehat{A}}$.

\begin{figure}[h]
\begin{center}
\includegraphics[width=0.68\textwidth]{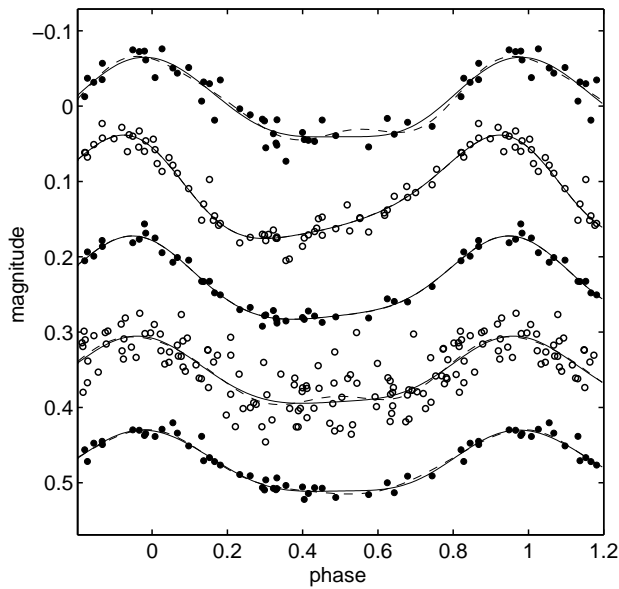}
\caption {Fitting of ``multicolour observational" data by advanced
PCA method (full line). Dashed lines represent the fitting by the
LSM.}\label{Fig2}
\end{center}
\end{figure}

\begin{figure}[h]
\begin{center}
\includegraphics[width=0.48\textwidth]{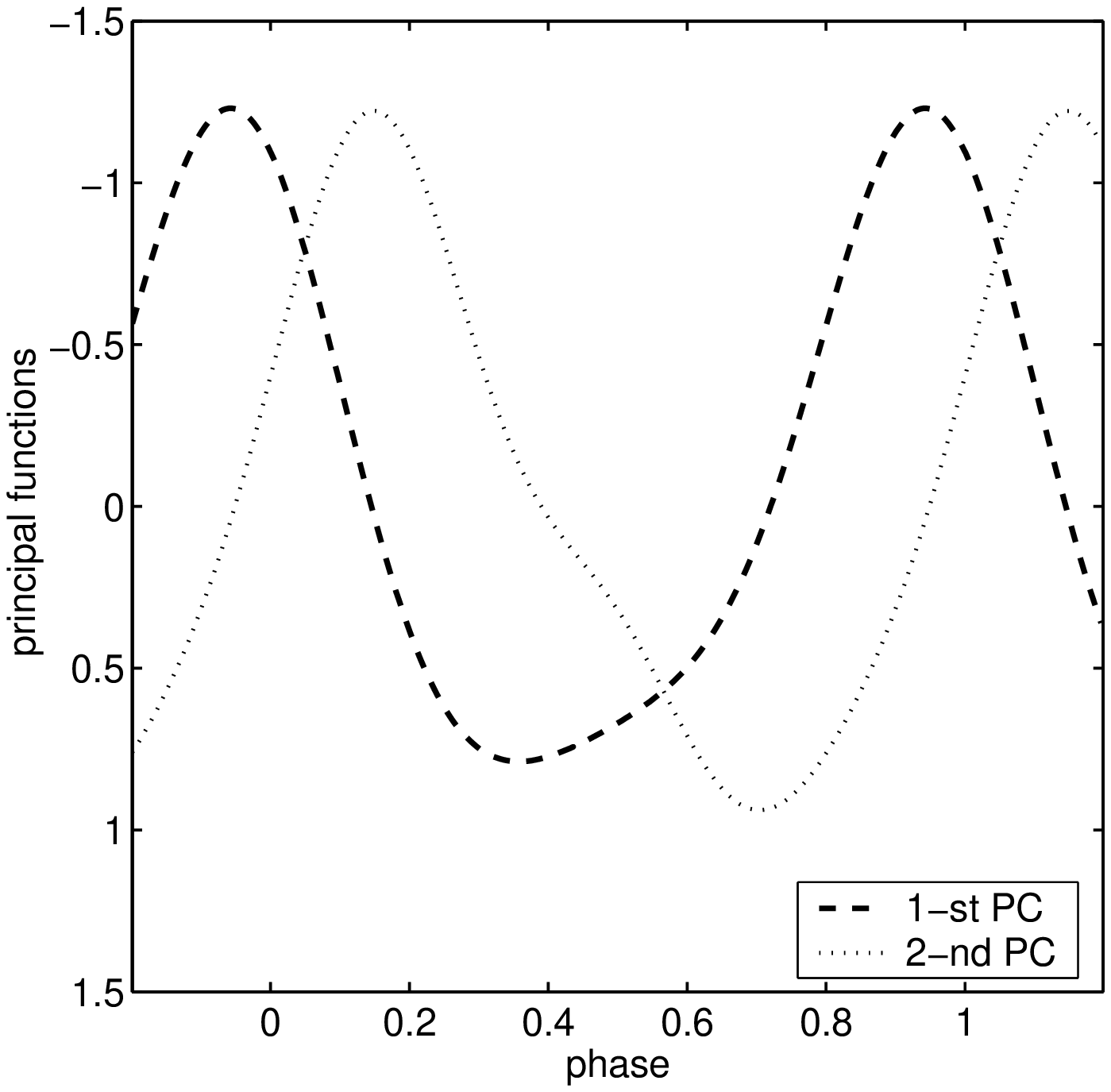}
\includegraphics[width=0.48\textwidth]{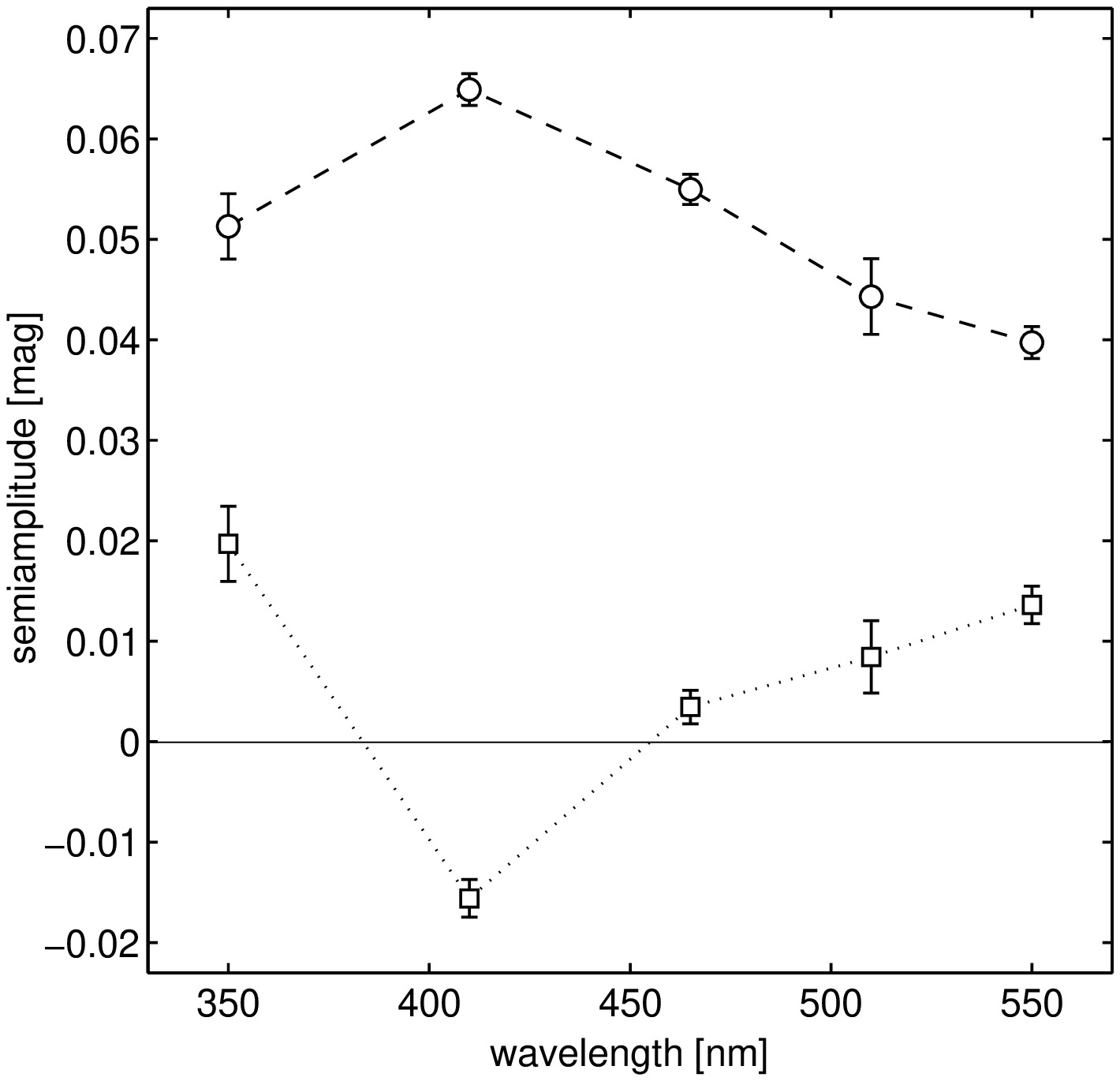}
\caption {(a) The courses of first two principal functions. (b) The
dependence of semiamplitudes on wavelength in nm for both principal
components. Dashed lines -- $1^{\mathrm{st}}$ principal component,
dotted lines -- $2^{\mathrm{nd}}$ principal component.} \label{Fig3}
\end{center}
\end{figure}

Nevertheless, we have to emphasize that the advanced PCA is not
identical with standard PCA or WPCA. APCA and PCA give very similar
but not the same results, smoothing matrix $\mathbf{\widehat{B}}$ is
not the duplicate of $\mathbf{\widehat{A}}$! The main reason
consists in the fact, that data treated by PCA have been centered to
their mean, while in the case of APCA we handle directly with the
found data without any centering. The difference is formulated in
the basic suggestion of APCA (\ref{Eq10}), which seems to us
physically more entitled than postulates of PCA. The correctness of
APCA method has been verified using several relevant statistical
tests and trials with simulated data.

We demonstrate the usage of APCA on synthetic photometric data
simulating light variability of the same model of the rotating CP
star. The Fig.~2 displays the phase diagram of multicolour light
variations:``observed" points are marked by full dots, the light
curves fitted by standard LSM technique are displayed by dashed
lines. The light curves found by the APCA are plotted by full lines
- they are indistinguishable from synthetic ones. The course of
first two principal curves are displayed in the Fig.~3a, the
dependencies of semiamplitudes versus both principal light curves
are plotted in the Fig.~3b. From the course of diagram of real
object we can gather on the nature of the light variability.

APCA can be used for reliable prediction of multicolour behaviour of
the object, the method is very apt for the light curves
quantification and classification \cite{MikGr2005},
\cite{MikZv2004}, for multicolour O-C measurements \cite{MikKr2007},
for the light ephemeris improvement \cite{MikGr2005}. APCA seems to
be a very efficient tool for the analysis of spectral variations and
radial velocity measurements \cite{KorMi2005}.

\section{Conclusions}

Principal Component Analysis and namely Advanced PCA proves to be an
universal, relatively simple method with an extremely versatile
extent of usage namely in the astronomical data (both photometric
and spectroscopic) processing and interpretation. Efficiency and
applicability of the PCA grows when we combine it with other
sophisticated methods of the data treatment as e.g. robust
regression, weighted LSM and wavelet analysis.

%acknowledgments
\vspace{5mm}

\noindent The author is very indebted for to Drs. Miloslav Zejda and
Jan Jan\'{i}k for careful reading of the manuscript and valuable
comments and suggestions. This investigation was supported by the
Grant Agency of the Czech Republic, grants No. 205/04/2063 and No.
205/06/0217.


\begin{thebibliography}{99}

\bibitem{Jack2003}
    Jackson, J.E., A User's Guide to Principal Components, Wiley,
    2003

\bibitem{Jol2004}
    Jolliffe, I.T., Principal Component Analysis, 2nd ed., Springer,
    2004

\bibitem{KorMi2005}
    Kor\v{c}\'akov\'a, D., Mikul\'a\v{s}ek, Z., Kawka, A., Kubat, J., Hornoch, K. et
    al., 2005, IBVS 5605, 1

\bibitem{MikGr2005}
    Mikul\'a\v{s}ek, Z., Gr\'af, T., 2005, Astrophys. Space Sci. \textbf{296}, 157

\bibitem{MikKr2007}
     Mikul\'a\v{s}ek, Z., Krti\v{c}ka, J., Zverko, J., \v{Z}i\v{z}\v{n}ovsk\'y,
     J., Jan\'{i}k, J., in the Proceedings of Active OB-Stars: Laboratories for
     Stellar $\&$ Circumstellar Physics, S. Stefl, S. Owocki and A. Okazaki eds.,
     ASP Conference Series Volumes, 2007, Vol. \textbf{361}

\bibitem{MikZv2004}
     Mikul\'a\v{s}ek, Z., Zverko, J., \v{Z}i\v{z}\v{n}ovsk\'y, J., Jan\'{i}k,
     J., in the Proceedings of The A-Star Puzzle, Poprad, Slovakia, J. Zverko,
     J. \v{Z}i\v{z}\v{n}ovsk\'y, S.J. Adelman, and W.W. Weiss eds.,
     IAU Symposium, No. 224. Cambridge, UK: Cambridge University Press, 2004, 657

\bibitem{MikRR2003}
     Mikul\'a\v{s}ek, Z., \v{Z}i\v{z}\v{n}ovsk\'y, J., Zverko, J.,
     Polosukhina, N.S., 2003, Contr. Astron. Obs. Skalnat\'e Pleso \textbf{33}, 29







\end{thebibliography}
\end{document}